# Emergent Properties of Terrorist Networks, Percolation and Social Narrative

Maurice Passman
Adaptive Risk Technology, Ltd.
London, UK
adaptiverisktechnology@gmail.com

Philip V. Fellman
American Military University
Charles Town, WV
shirogitsune99@yahoo.com

**Abstract**

In this paper, we have initiated an attempt to develop and understand the driving mechanisms that underlies fourth-generation warfare (4GW). We have undertaken this from a perspective of endeavoring to understand the drivers of these events (i.e. the 'Physics') from a Complexity perspective by using a threshold-type percolation model. We propose to integrate this strategic level model with tactical level Big Data, behavioral, statistical projections via a 'fractal' operational level model and to construct a hierarchical framework that allows dynamic prediction. Our initial study concentrates on this strategic level, i.e. a percolation model.

Our main conclusion from this initial study is that extremist terrorist events are not solely driven by the size of a supporting population within a socio-geographical location but rather a combination of ideological factors that also depends upon the involvement of the host population. This involvement, through the social, political and psychological fabric of society, not only contributes to the active participation of terrorists within society but also directly contributes to and increases the likelihood of the occurrence of terrorist events.

Our calculations demonstrate the links between Islamic extremist terrorist events, the ideologies within the Muslim and non-Muslim population that facilitates these terrorist events (such as Anti-Zionism) and anti-Semitic occurrences of violence against the Jewish population.

In a future paper, we hope to extend the work undertaken to construct a predictive model and extend our calculations to other forms of terrorism such as Right Wing fundamentalist terrorist events within the USA.

**Emergent Properties of Terrorist Networks, Percolation and Social Narrative**

*God is Greatest, Death to America, Death to Israel, Curse on the Jews, Victory to Islam*

Motto on the Houthi group's flag

*Always blame the Jews for everything*

Newsweek, 12/13/17, Report on Neo-Nazis in the USA

## 1.0 Introduction

Statistics is the science of mathematically examining data. From this data, we seek patterns and correlations that hopefully explain what is going on within the data. Predictions made from statistics can be compared to driving a car whilst looking out of the rear window. Causality is implicit and follows what logicians call the fallacy of affirming the consequent. We observe b following a and then intuit that a causes b. In statistics, this inference is supported by the process of mathematical correlation between observations. We do not observe causality directly but rather build our theories on correlations between observed behavioral regularities. This 'statistical' methodology lies behind the current 'Big Data' processes in that data volume and computing power allows individualization of behavior and rapid (short time period) predictions. Physics, on the other hand, takes these observed behavioral regularities and derives theories of causation which are designed to make accurate and informed predictions. The distinction between the two sciences (physics and statistics) may be subtle but is profound.

In this paper, we wish to develop and understand the physics that underlies terrorist events. These events may be seen as part of an overall new type of warfare where nation states have lost the monopoly of war. This new generation of warfare, labeled the fourth generation (4GW) [1], is that of wars undertaken by cultures against States. This type of warfare is, by definition, asymmetrical and defeat of the nation state my not necessarily be military but rather political. This paper concentrates on European Islamic terrorist events and UK anti-semitism as an ideological indicator simply due to the paucity of other data sources. The main conclusions from this initial study may be controversial but they are also far reaching: non-active dissuasion of terrorist ideologies (which we call 'passive support') populations not only directly contributes to the active participation of terrorists within society but also increases the likelihood of the occurrence of terrorist events. These conclusions have both social and political implications. The drivers for our mathematical framework are founded not upon facts (the 'truth') but rather narrative[1]: it is a set of beliefs within a population or population sub-set (an ideology) that allows the construction of an ideological sea within which terrorists can swim. This entails that ideological beliefs, whether they are false or not, become the 'truth' as they are axiomatic[2]. We see this ideological effect across all sides of the political spectrum particularly in the UK. Examples of this are:
- The inability of the UK Labour Party to come to grips with its perceived institutionalized anti-

---

[1] For example, the 'belief' rather than the 'fact' that 'Jews control the media'. The concept that a population is constrained by a belief system maintained by the process of capitalization (in that non-material concepts as well as material objects may be commodities in the Marxian sense) was propounded by Guy Debord in his 1967 book 'The Society of the Spectacle'. An event or series of events are therefore then required to occur so that the 'truth' may be seen (in Debord's language to create a 'Situation'). A more recent example of this is Neo having to choose the red or blue pill in the Matrix film.

[2] See for example https://www.telegraph.co.uk/education/2018/07/02/labour-responsible-rise-softcore-holocaust-denial-dr-deborah/

Semitism[3] (a political belief system of the Left)[4];
- The inability of elements within the UK Muslim community to acknowledge or actively dissuade Muslim anti-Semitism (a religious belief system)[5]
- The rise of popularism within Europe and the UK[6] (a political belief system of the Right);
- The rise of casual, every day, antisemitism in the UK[7].

In examining the Physics of these events, we seek to develop a theory which can lead to the efficient interdiction of terrorist behaviors. Our conclusion, therefore, is to propose to construct a hierarchical model that benefits from the advantages of both Complexity Science and Big Data Statistical methodology. Group behavior (i.e. the predictive modeling of 'the whole is greater than the sum of the parts') at a strategic level will be modeled by a Complexity (Percolation) type model, whilst individualization modeling ('tactical level', granular, data intensive Big Data model) would be applied in a ground up manner. Thus, the strategic and tactical would be able to validate each other dynamically.

On 20 July 2015, at Ninestiles School in Birmingham, the then UK Prime Minister, David Cameron, set out his policy plans to address extremism and Islamic extremism [2]. This policy was part of the UK Government's detailed response to ISIL, Iraq and Syria. Specifically, the Prime Minister gave four main reasons/causes for segments of the UK populace being drawn to Islamic terrorism:

- Like any extreme doctrine, *Jihadism* can seem energizing, especially to young people;
- The process of extremism is step-wise and starts with *radicalization*. When you look in detail at the backgrounds of those convicted of terrorist offences, many of them were first influenced by what some would call non-violent extremists;
- Extremist ideology drives and sets the terms of the public and political debate;
- With respect to the question of identity, there are many people born and raised in the UK who do not truly identify with the UK. With respect to Islamic extremism, these individuals identify themselves first with a particular extremist creed, second with a particular religious and ethnic group and see their UK citizenship primarily as an accident of location.[8]

The Prime Minister then went on to outline a number of policy directives that were aimed at countering the above four causes of terrorism:

- Confronting, head on, extremist ideology, particularly by taking its component parts to pieces - especially the cultish worldview and the conspiracy theories of jihadism;
- Tackling both parts of the extremist creed – the non-violent and the violent. This means confronting groups and organizations that may not advocate violence – but which do promote other parts of the extremist narrative;
- Embolden different voices within the Muslim community and actively encourage the reforming and moderate Muslim voices;
- The fourth and final part of the strategy is to build a more cohesive society, so that more people feel a part of it and are therefore less vulnerable to extremism and anti-social behavior.

---

[3] See https://blogs.spectator.co.uk/2018/04/jeremy-corbyn-and-the-far-lefts-anti-semitism-doublespeak/
[4] See https://www.thejc.com/comment/analysis/jeremy-corbyn-labour-definition-antisemitism-1.466626
[5] See https://musliminstitute.org/freethinking/culture/sorry-truth-virus-anti-semitism-has-infected-british-muslim-community
[6] See https://en.wikipedia.org/wiki/Right-wing_populism
[7] See https://www.theguardian.com/commentisfree/2016/may/03/everyday-antisemitism-britain-prejudice
[8] Bernard Lewis (1998) in The M*ultiple Identities of the Middle East*, explains that within Islamic society, identity is conceived of in a fundamentally different fashion than it is in the West. He explains that historically, for example, the Ottoman Empire saw itself as the geographic successor to the Eastern Roman Empire, and still made references to this in official documents as late as the 19th century. In terms of identity, religious and tribal affiliations have virtually always superseded geographic location in Islamic culture. [3]

It is always advisable that governmental policy be supported by sound scientific discourse. The aim of this paper therefore is to use the above four-point policy framework to provide a foundation by which we may build a dynamic model that describes, and hopefully goes some way towards predicting the occurrence of terrorist events. The process will follow studies undertaken with some success previously by the authors for attrition warfare by constructing a hierarchy of meta-models (i.e. building of an aggregated system of models giving greater and greater granularity) using the tools of Complexity Science: fractals, self-organization, self-similarity and scale-free systems [4]. Note that within this paper, we work within a Eurocentric context; there are no reasons for not extending this work to other geo-social and geo-spatial areas. However, the initial paucity of data, and the fact that a previous study was based upon Spanish data (notably reference [14]) restrained us from extending our scope in the present paper.

In previous work [4,5] we have presented evidence that suggests that further development, with agent based models, together with mathematical techniques such as dimensional analysis and graph theory, can be linked to historical data and fractal fingerprint behavior (i.e. a meta model hierarchy). In this prior work, we presented the idea that Statistical Thermodynamics could be used to examine agent based models of the geo-social process that is 'Combat' (i.e. where humans come together within a social and geographical context called 'War' and attempt to kill each other).

This set of insights naturally led us onto investigating percolation theory, critical systems, Barenblatt's scaling and dimensional analysis methods [6], renormalization theory [7] and Per Bak's self-organized criticality (SOC)[8]. Employing these methods eventually resulted in the derivation of a 'Universal Law' for attrition warfare, namely, the Fractal Attrition Equation [4]. Our present aim therefore is to develop a similar set of descriptors for terrorism. Predictive models are often touted as a way to test ideas, concepts or new theories. While those things can be simulated by an algorithmic model, the results from such a model do not 'prove' anything. What this kind of model can do, however, is to shine a light on which interactions and relationships are important. Simple qualitative or descriptive studies cannot adequately reveal these features nor illumine their relationships. The current paper has been designed with the purpose of providing a springboard by which we hope to gain insight into the dynamics that drive terrorist events and thus reveal the predictive drivers. We are thus, currently developing the tools to examine, model and predict a wider range of behavioral and associated processes. We hope to publish these results in a following paper.

Our first question was how were we to construct a high-level model that described the human environment in which terrorists evolve, i.e. the social and geographical network space of the jihad. Fortunately, there exists a model system that can be established as a high-level starting point: Serge Galam's percolation model [9]. Percolation is concerned with connectedness. For an example, we may take a board from the game of "Go". If two stones are on adjacent sites they are said to be connected. After randomly placing stones on the board we get to a stage where it is possible to trace a connection from one side of the board to the other: this is where it is said that we have percolation. So, at the precise point where a connected path appears connecting all of the islands, percolation is said to occur. Many natural phenomena act as if they are made of random 'islands' and under certain conditions, over time, one macroscopic continent emerges. Connectedness in a geo-social context is called social permeability. Social permeability also describes the physical pathways that nodes (individual extremists) of the extremist networks can establish and use to move along freely and safely. In the Galam model, this freedom of movement is due to what Galam calls 'passive supporters' of the extremist cause [10]. Note that permeability is satisfied without necessarily having direct physical contact or even communication between the extremist nodes and the mass of passive supporters. Galam defines passive supporters of the jihadist cause as 'normal' people who do not necessarily express their position explicitly; often this is an individual attitude associated with a personal opinion. It does not need to be explicitly so. They are unnoticeable, and most of them reject the violent aspect of the jihadist action. They only share their cause in part. The degree of permeability of the social context determines the limits of the geo-social space open to jihadist actions. This is a different form of analysis than the social network analysis which we used, for example, to analyze the 9/11 Hijackers' network insofar as it does not depend upon Watts-Strogatz small world network relationships or a Hamiltonian path [11][12].

## 2.0 The Galam Percolation Model of Terrorism

Our starting point was Serge Galam's percolation model [9]. Galam models terrorism as a function of the percolation of active terrorists across a complex landscape of passive cells which either support (allow percolation) or do not support (deny) percolation to the terrorist. This landscape is represented by a grid whereby for a single individual to move across the lattice he must pass from one cell to another, via the intermediate adjacent cells. Let us consider a terrorist node that wishes to move safely from its nucleus to another social space in the lattice. In Galam's model, the relative number of passive supporters of the terrorist cause with respect to the total population in a given region is compared to the critical percolation threshold for the said region. This comparison is what determines the effective scope of the terrorist threat. As soon as the density of passive supporters of the terrorist cause rises above the percolation threshold, the entire territory is under threat of terrorist action. In the context shown here, it is not sufficient to neutralize the nodes of the terrorist network to ensure that a specific area remains free from threat. Once the terrorist network has been stopped, another one will, after a time, take its place and so conserve intact the possibility of a new, immediate attack in all the social-geographical space accessible to its members via the percolant pathways [13]. The pathways through the region which have been established by the passive supporters of the terrorist cause remain intact unless and until specific action is taken to dismantle them or otherwise lower their density below the percolation threshold. From this point of view, passive supporters of the terrorist cause are the weak ties of terrorist networks, but they are also critical enablers of terrorist attacks.

Galam argues that in order for a terrorist to act they must find a path within a safe social space. Thus, the determination of the terrorist range of destruction is not merely based upon the terrorist net itself but on the geographical and social permeability of its members; much in the way that Mao stated that "A revolutionary activist must be like a fish in water while in the popular masses". In the Galam model, the relative number of passive supporters is compared with the total population of a given region and a critical percolation threshold calculated. Therefore, terrorist groups and networks are seen to interact and move through the social geographical 'phase space' made accessible by non-passive support. The adaptivity and power of jihadist networks hence lies in their ability to navigate the landscape of passive supporters. In Galam's percolation model, for a given territory, the distribution and size of aggregate spaces yielded by its passive supporters directly determines the range of terrorist action. It is the relative value of the passive support of the population $p$ compared with the value of the critical point $p_c$ of the corresponding space that determines the effective amplitude of the terrorist threat. In other words, the passive support of the population $p$ yields a threshold value, that when compared with the value of the critical point $p_c$ of the corresponding space, determines whether the line has been crossed and a terrorist event will occur. If the passive support is larger than the percolation threshold then all territory falls under terrorist threat. We can calculate the value of the critical point critical point $p_c$. A universal formula holds for site percolation where:

$$p_c = a[(d-1)(q-1)]^{-b} \qquad (1)$$

where $q$ is the connectivity of the network (the average number of immediate neighbors of a site), $d$ is the dimension of the space within which the network is embedded and with $a = 1.2868$ and $b = 0.6160$. (See the following diagram from Galam's original study.)

**Figure 1: Example of the relationships between Connectivity, Dimension & Threshold within the Universal Site Percolation Formula (from [9])**

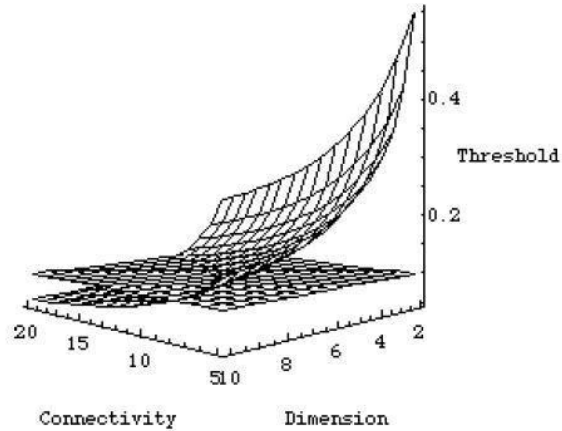

Figure 5: Representation of the Galam-Mauger universal formula [17] for all the thresholds of percolation as function of connectivity and dimension. The formula writes $p_c = a[(d-1)(q-1)]^{-b}$ where $d$ is dimension, $q$ connectivity, $a = 1.2868$ and $b = 0.6160$.

In the Galam model, much of the enabling behavior for terrorism simply requires a landscape with a minimum number of passive supporters. In most cases a small number of passive supporters is sufficient to create a landscape over which terrorist activity can percolate. Terrorist deployment thus obeys a universal scheme of activity with two phases, a percolating phase and a non-percolating phase. The only difference from one form of terrorism to another is the scale on which passive supporters are spread and the geographical area on which percolation may take place. Obviously if the change of scale does not change the nature of the terrorist phenomenon, it modifies in a substantial manner the number of threatened people. This change in scope is clearly not a negligible difference. It is the case of September 11, which while revealing for the first time the existence of a world-wide percolation also showed simultaneously that from now on, the entire world's population could be in danger.

Of critical importance is Galam's demonstration that phase transition from landscapes where percolation is impossible to landscapes where percolation is possible requires only a small addition of passive supporters. He argues that because of this, overwhelming military presence will not create a solution to terrorism, and that a more fruitful approach may be to reduce the social dimensionality of the percolation space. Galam suggests that for a social application of percolation, the network connectivity could be of the order of 15 and the dimension be 2 (i.e. the 2 dimensional 'flat' surface of the earth). For these suggested values, $p_c = 0.25$. In other words, a density of 25% of passive supporters - this means that 25% of the earth's population are passive supporters of terrorism. Galam goes on to conclude that this percentage is 'totally out of context': he keeps $q = 15$ and changes $d$ to 15. This yields a value of 6% which he deems to be a much closer representation of the real value (i.e. 6% of the world's population are passive supporters). Galam's reasoning is that we are actually dealing with a geographical-social lattice, so the dimension $d$ is the sum of the physical dimension $d_F = 2$ and a memetic dimension $d_M$, where $d = d_F + d_M$. We shall look at the 25% value presently. Miralles Canals [13] suggests the following list of 10 memetic dimensions:

**Figure 2. Memetic values chosen by Miralles Canals given in [14]**

Table 6
Memetic table associated to the passive supporters of the jihadist cause in Spain.

| Type | Meme |
|---|---|
| Defensive jihad | Individual obligation of salvation |
| Defensive jihad | Territorial revindication of all the Iberian Peninsula |
| Defensive jihad | Revindication of the cultural superiority of Islam |
| Devensive jihad | Fulfilment of direct prophecy attributed to Mohammed |
| Defensive jihad | Punishment for apostasy |
| Non-defensive jihad | Ecologist need, "Islamic culture green" |
| Non-defensive jihad | Hatred of the capitalist system and of the West |
| Non-defensive jihad | Hatred of Jews and the Nation of Israel |
| Non-defensive jihad | Revindication of Spanish territory by Moslem Nations |
| Non defensive jihad | Hatred of Catholicism in Spain |

Note should be taken that it is the memes – the narrative or the ideology that these memes are indicators for – that drive the process. In the political dimension this means that the extremists drive the process. But they require an environment, oxygen, to drive the process, the fire. Therefore, there is a feedback structure between this extremist, political group, and the population that moves the system (or not) to the critical point[9]. In order to study the actual possible relationships that drive the universal percolation formula we examined ethnographic data from Spain and the UK. The Spanish calculations are based on Miralles Canals paper of 2009 [14] (we have also corrected some of the numerical errors in this paper). Miralles Canals' original calculations are shown below for reference:

**Figure 4. Miralles Canals' original calculation results as given in [14]**

Table 5
Social and political space of the jihad: an estimate for the kingdom of Spain.

| AC | TP | MP | NJW | PSJ | p $10^{-5}$ |
|---|---|---|---|---|---|
| Catalonia | 7 354 441 | 279 027 | 9–36 | 37 111–93 000 | 505–1265 |
| Madrid | 6 251 876 | 196 689 | 7–26 | 26 160–65 556 | 356–891 |
| Andalucia | 8 177 805 | 184 430 | 6–24 | 24 529–61 471 | 334–836 |
| Valencia | 5 016 348 | 130 471 | 4–1 | 17 353–43 486 | 236–591 |
| Murcia | 1 424 063 | 63 040 | 2–8 | 8 384–21 011 | 114–286 |
| Canary Isles | 2 070 465 | 54 636 | 2–7 | 7 267–18 210 | 99–248 |
| Melilla | 71 339 | 34 397 | 1–5 | 4 575–11 465 | 62–156 |
| Castilla-La Mancha | 2 038 956 | 32 960 | 1–4 | 4 384–10 986 | 60–149 |
| Aragón | 1 325 272 | 30 982 | 1–4 | 4 121–10 327 | 56–140 |
| Ceuta | 77 320 | 30 537 | 1–4 | 4 061–10 178 | 55–138 |
| Balears | 1 071 221 | 25 859 | 1–3 | 3 439–8 619 | 47–117 |
| Castilla y León | 2 553 301 | 17 336 | 1–2 | 2 306–5 778 | 31–79 |
| Basque Lands | 2 155 546 | 16 608 | 0–1 | 2 209–5 536 | 30–75 |
| Extremadura | 1 095 894 | 15 536 | 1–2 | 2 066–5 178 | 28–70 |
| Navarra | 619 114 | 10 884 | 0–1 | 1 448–3 628 | 20–49 |
| La Rioja | 317 020 | 10 373 | 0–0 | 1 380–3 457 | 19–47 |
| Galicia | 2 783 100 | 6 079 | 0–0 | 809–2 026 | 11–28 |
| Asturias | 1 079 215 | 2 731 | 0–0 | 363–910 | 5–12 |
| Cantabria | 581 215 | 2 179 | 0–0 | 290–726 | 4–10 |

AC = Autonomous Community, TP = Total population, MP = Moslem population, NJW = Nodes of jihadist networks, PSJ = Passive supporters of the jihadist cause, p = Relative number of supporters of the jihadist cause with respect to the total population.

Miralles Canals assumes the following values taken via an application of the Clarke Layers model [14]. The Clarke layers model is simply an estimate of the classification of extremists within the world Muslim population:

---

[9] As an example of our thinking, Hitler won the German federal election of July 1932 with a 37.27% vote. Thus, the ideological 'meme's' that fed the political process to the critical point (the 'win') were 'pumping' or 'stoking' the percolation mechanism. Hitler maneuvered and choreographed his ideological outlook depending upon the feedback from his audience i.e. the collective consciousness of the people that supported him. It was the 60 or so percent of the rest of the population that *allowed* this process to reach critical point.

**Figure 5. Clark layer values chosen by Miralles Canals given in [14]**

| World Muslim Population | $1 \times 10^8$ |
|---|---|
| Jihadist Supporters | 200 to 500 x $10^6$ = <br> 13.33 to 33.33% |
| Jihadist Groups | 50 to 200 x $10^3$ = <br> 0.0033 to 0.0133% |
| Active Terrorists | 200 to 500 x $10^3$ = <br> 0.000027 to 0.000133% |

Applying the Clarke Layers to the Muslim demographic population values given in the Miralles Canals paper we obtain the following results:

**Figure 6. Social and Political space for Spain as calculated by the Authors**

| Autonomous Community AC | Total Population (TP) | Moslem Population (MP) | % Ratio MP/TP | NJW Lower Bound | NJW Upper Bound | PSJ Lower Bound | PSJ Upper Bound | p 10^-5 Lower Bound | p 10^-5 Upper Bound |
|---|---|---|---|---|---|---|---|---|---|
| Catalonia | 7354441 | 279027 | 3.8% | 9 | 36 | 37194 | 93000 | 506 | 1265 |
| Madrid | 6251876 | 196689 | 3% | 6 | 26 | 26219 | 65556 | 419 | 1049 |
| Andalucía | 8177805 | 184430 | 2% | 6 | 24 | 24585 | 61471 | 301 | 752 |
| Valencia | 5016348 | 130471 | 3% | 4 | 17 | 17392 | 43486 | 347 | 867 |
| Murcia | 1424063 | 63040 | 4% | 2 | 8 | 8403 | 21011 | 590 | 1475 |
| Canary Isles | 2070465 | 54636 | 3% | 2 | 7 | 7283 | 18210 | 352 | 880 |
| Melilla | 71339 | 34397 | 48% | 1 | 4 | 4585 | 11465 | 6427 | 16070 |
| Castilla-La Mancha | 2038956 | 32960 | 2% | 1 | 4 | 4394 | 10986 | 215 | 539 |
| Aragon | 1325272 | 30982 | 2% | 1 | 4 | 4130 | 10326 | 312 | 779 |
| Ceuta | 77320 | 30537 | 39% | 1 | 4 | 4071 | 10178 | 5265 | 13163 |
| Balears | 1071221 | 25859 | 2% | 1 | 3 | 3447 | 8619 | 322 | 805 |
| Castilla y Leon | 2553301 | 17336 | 1% | 1 | 2 | 2311 | 5778 | 91 | 226 |
| Basque Lands | 2155546 | 16608 | 1% | 1 | 2 | 2214 | 5535 | 103 | 257 |
| Extremadura | 1095894 | 15536 | 1% | 1 | 2 | 2071 | 5178 | 189 | 473 |
| Navarra | 619114 | 10884 | 2% | 0 | 1 | 1451 | 3628 | 234 | 586 |
| La Rioja | 317020 | 10373 | 3% | 0 | 1 | 1383 | 3457 | 436 | 1091 |
| Galicia | 2783100 | 6070 | 0% | 0 | 1 | 809 | 2023 | 29 | 73 |
| Asturias | 1079215 | 2731 | 0% | 0 | 0 | 364 | 910 | 34 | 84 |
| Cantabria | 581215 | 2179 | 0% | 0 | 0 | 290 | 726 | 50 | 125 |

**Figure 7. Bounds for Social and Political space for Spain as calculated by the Authors**

| Autonomous Community AC | PSJ % of TP Lower Bound | PSJ % of TP Upper Bound | p Upper Bound | p for Upper Bound as a % |
|---|---|---|---|---|
| Catalonia | 1 | 1 | 0.01 | 1.26 |
| Madrid | 0 | 1 | 0.01 | 1.05 |
| Andalucía | 0 | 1 | 0.01 | 0.75 |
| Valencia | 0 | 1 | 0.01 | 0.87 |
| Murcia | 1 | 1 | 0.01 | 1.48 |
| Canary Isles | 0 | 1 | 0.01 | 0.88 |
| Melilla | 6 | 16 | 0.16 | 16.07 |
| Castilla-La Mancha | 0 | 1 | 0.01 | 0.54 |
| Aragon | 0 | 1 | 0.01 | 0.78 |
| Ceuta | 5 | 13 | 0.13 | 13.16 |
| Balears | 0 | 1 | 0.01 | 0.80 |
| Castilla y Leon | 0 | 0 | 0.00 | 0.23 |
| Basque Lands | 0 | 0 | 0.00 | 0.26 |
| Extremadura | 0 | 0 | 0.00 | 0.47 |
| Navarra | 0 | 1 | 0.01 | 0.59 |
| La Rioja | 0 | 1 | 0.01 | 1.09 |
| Galicia | 0 | 0 | 0.00 | 0.07 |
| Asturias | 0 | 0 | 0.00 | 0.08 |
| Cantabria | 0 | 0 | 0.00 | 0.12 |

The recalculated results are significant in a number of ways. First, it should be noted that the passive supporter values for Ceuta and Melilla are much larger than those for Catalonia, Madrid and Andalucía. According to these values, therefore, we would expect a greater probability of terrorist events for Ceuta and Melilla than for Catalonia, Madrid and Andalucía (all of which have larger Muslim populations but smaller populations in proportion to the non-Muslim population). But this has not been the case in reality. Note also that the number of terrorist nodes is fewer for Ceuta and Melilla than it is for Catalonia, Madrid and Andalucía.

One explanation for this disparity can be attributed to the method by which passive supporter numbers are calculated in [14]: the values are calculated from the numerical values of the Muslim population within an Autonomous Community; there are *no* non-Muslim contributors to the passive supporter values in the above tables. The $p$ value for Ceuta is around 13%, in other words 13% of the Muslim population of Ceuta are calculated as passive supporters of terrorism, whilst the $p$ value for Madrid (where the bombings actually occurred is around 1%). From a conceptual viewpoint, there is no reason why passive supporters should be drawn solely from the Muslim population. Our conclusion, therefore, is that passive supporters from the non-Muslim population (signified as a percentage of the *non-Muslim* population within an Autonomous Community) should be contributing to the $p$ value. Do we see something similar in other data? We would have liked to have included French data in our study (due to the increased occurrences of terrorism in that country) but for historical reasons France does not collect data on religious demographics. We have therefore used UK data [6] (from the 2011 census published in 2015) to undertake a $p$ value calculation similar to the Spanish examples:

**Figure 8. Social and Political space for UK as calculated by the Authors**

| Region | Total Population (TP) | Moslem Population (MP) | % Ratio MP/TP | NJW Lower Bound | NJW Upper Bound | PSJ Lower Bound | PSJ Upper Bound | p $10^{-5}$ Lower Bound | p $10^{-5}$ Upper Bound |
|---|---|---|---|---|---|---|---|---|---|
| London | 8173941 | 1012823 | 12.4% | 33 | 132 | 135009 | 337574 | 1652 | 4130 |
| West Mids | 5601847 | 376152 | 6.7% | 12 | 49 | 50141 | 125371 | 895 | 2238 |
| North West | 7052177 | 356458 | 5.1% | 12 | 46 | 47516 | 118807 | 674 | 1685 |
| Yorks and Humber | 5283733 | 326050 | 6.2% | 11 | 42 | 43462 | 108672 | 823 | 2057 |
| South East | 8634750 | 201651 | 2.3% | 7 | 26 | 26880 | 67210 | 311 | 778 |
| East of England | 5846965 | 148341 | 2.5% | 5 | 19 | 19774 | 49442 | 338 | 846 |
| East Mids | 4533222 | 140649 | 3.1% | 5 | 18 | 18749 | 46878 | 414 | 1034 |
| South West | 5288935 | 51228 | 1.0% | 2 | 7 | 6829 | 17074 | 129 | 323 |
| North East | 2596886 | 46764 | 1.8% | 2 | 6 | 6234 | 15586 | 240 | 600 |

**Figure 9. Bounds for Social and Political space for UK as calculated by the Authors**

| Region | PSJ % of TP Lower Bound | PSJ % of TP Upper Bound | p Upper Bound | p for Upper Bound as a % |
|---|---|---|---|---|
| London | 2 | 4 | 0.04 | 4.13 |
| West Mids | 1 | 2 | 0.02 | 2.24 |
| North West | 1 | 2 | 0.02 | 1.68 |
| Yorks and Humber | 1 | 2 | 0.02 | 2.06 |
| South East | 0 | 1 | 0.01 | 0.78 |
| East of England | 0 | 1 | 0.01 | 0.85 |
| East Mids | 0 | 1 | 0.01 | 1.03 |
| South West | 0 | 0 | 0.00 | 0.32 |
| North East | 0 | 1 | 0.01 | 0.60 |

From the above calculations, it can be seen that the UK has a different demographic from Spain insofar as there are no regions within the UK that mirror the high Muslim population density of Ceuta and Melilla. Secondly, Islamic terrorist events have occurred within London (and this corresponds to London having the highest $p$ value) and not other regions, so we must look for other data to examine if there are any visible patterns concerning non-Muslim population effects within the UK. Whilst the West (as personified by America and its ally the UK) is often the target of jihadist attacks, a number of attacks have also targeted the Jewish community and have often done so simultaneously with other terrorist events. One argument behind the reason the Jewish community often is targeted by Muslim extremists is that Jews provide an existential threat to extremist Muslims: the very existence of Jews is a form of proof that Islam is based on a false premise (i.e., if Jews were alive at the time of Mohammed, then why did they not become Muslims? A characteristic of some Islamic terrorist events is that they contain both the language and the actions of humiliation and killing of Jews which makes up part of the 'proof' that Islam is God's 'Final Testament'). It is not surprising, therefore, that Jewish targets were struck as part of the Charlie Hebdo events in Paris. Our next question therefore was: what is the pattern of violence that the UK Jewish population was subjected to that possibly mirrors the threshold calculations in the above tables? The UK demographics that include Jewish and Muslim data are given in the following table:

**Figure 10. Population demographics as given by the 2011 UK Census [6]**

| Location | Population: All | Population: Jewish | Population: Muslim | Muslim Ratio to Jews | % Muslims in Population |
|---|---|---|---|---|---|
| England & Wales | 56075912 | 263346 | 2706066 | 10:1 | 5 |
| England | 53012456 | 261282 | 2660116 | 10:1 | 5 |
| London | 8173941 | 148602 | 1012823 | 7:1 | 12 |
| West Mids | 5601847 | 4621 | 376152 | 81:1 | 7 |
| North West | 7052177 | 30417 | 356458 | 12:1 | 5 |
| Yorks and Humber | 5283733 | 9929 | 326050 | 33:1 | 6 |
| South East | 8634750 | 17761 | 201651 | 11:1 | 2 |
| East of England | 5846965 | 34830 | 148341 | 4:1 | 3 |
| East Mids | 4533222 | 4254 | 140649 | 33:1 | 3 |
| South West | 5288935 | 6365 | 51228 | 8:1 | 1 |
| North East | 2596886 | 4503 | 46764 | 10:1 | 2 |
| Wales | 3063456 | 2064 | 45950 | 22:1 | 1 |

The CST Report for 2013 [16] states that over three-quarters of the total anti-Semitic incidents (of all types) recorded in 2013 took place in Greater London and Greater Manchester (the North West Region), the two largest Jewish communities in the UK. Within Greater London, the borough where the highest number of anti-Semitic incidents was recorded was Barnet. Barnet has the largest Jewish community of any local authority in the UK. Anti-Semitic attacks are greatest in number not only in areas of high Jewish population but also of regions of high non-Jewish and non-Muslim population but the CST does not give a breakdown of the demographics behind the incidents (e.g. far right Anti-Semitism verses any other type).

From the CST Report, we find that the general breakdown of types of anti-Semitic incidents are as follows: Anti-Semitic incidents in Greater Manchester are more likely to involve random street racism – what might be called anti-Semitic hooliganism – against individual Jews. Ideologically motivated Anti-Semitism tends to be concentrated in Greater London where most of the Jewish community's leadership bodies and public figures are based and where there is the greatest population of Jews. For example, 55% of anti-Semitic incidents recorded by CST in Greater Manchester targeted individual Jews in public, compared to 25% of the incidents recorded in Greater London. Similarly, 21% of incidents recorded in Greater London targeted Jewish organizations, events or communal leaders, compared to just 2 % of the incidents in Greater Manchester. However, 'Abusive behavior' was more common in London by a factor of 1.8:1 but reported assault was less in London than Manchester (by a factor of 1:1.4).

To summarize, Anti-Semitic incidents are greatest where there is the greatest concentration of Jews in a Region (and therefore the greatest visibility). London has a larger Jewish community than Manchester. Note that the proportion of Jews to Muslims does not seem to be a factor here. The West Midlands, where the ratio of Muslims to Jews is 81:1, has few recorded incidents of Anti-Semitism. The greatest levels of anti-Semitic attacks are in regions where there is the largest overall population (London and the North West). The level of anti-Semitic violence is greater for Manchester than it is in London possibly because of the visibility of the Jewish population combined with a 12:1 ratio of Muslims to Jews. Our next question therefore is "If the level of Anti-Semitism and by implication Islamic extremist 'events' is linked to the Jewish population, the Muslim population and non-Muslim passive supporters, then is there something that links all three?" The answer is a resounding "Yes" and the thing which most profoundly links all three is Israel.

What are the reasons for non-Muslims to be passive supporters of terrorism? First, we should look at why non-Muslims would be antagonistic (passive or non-passive) towards the State of Israel (as opposed being anti-Semitic). The ideological causes of antagonism fall into three general areas:

- The belief in civil rights over security - the argument that the humiliation that has been experienced by the Palestinians cannot be justified by continued Israeli fears of terrorist attacks. In its more extreme forms, this argument can be extended to the idea that Israeli defense measures are actually "aggression" against the Palestinians and against other states in the region.
- The belief that it is morally correct to support the (perceived) underdog. This position is justified by the argument that someone's material position is mostly a function of their conditions rather than their choices. In this context, the argument further goes that the Palestinians are much poorer than the Israelis because of the way Israel controls and suffocates the Palestinian economy, which in turn is what leads Palestinians to radicalism. This argument is in part a recapitulation of nationalist arguments made throughout the twentieth century and often emerges in one form or another as the thesis that rich nations are the cause of poverty and misfortune in poor nations.
- The "politically correct" notion that the conflict between Israel, the Palestinians and Israel's Arab neighbors is one between a white people and a non-white people, and it is therefore morally correct to support the non-white people. In this context, however incorrectly, Israelis are identified as white and Palestinians as non-white.

However, criticism of Israel is not necessarily Anti-Semitism. As a result, the question arises as to how the level of Anti-Semitism and (by implication) Islamic extremist 'events' is connected to (a) the size of the Jewish population, (b) the size of the Muslim population and (c) the size of the non-Muslim passive supporters. This question can be particularly difficult to unravel. Part of the answer lies in the fact that there is only one Jewish state and that state is irrevocably tied up with the Jewish population outside of Israel. The reason for this link is not necessarily 19th Century type nationalism (or the twentieth century type nationalism referenced above) but rather, that Israel was seen, and still is, by Jewish victims as a possible haven from anti-Semitic violence[10]. However, the continued violence within the area currently occupied by the State of Israel against Jews both before and after the founding of the State of Israel has meant that Israel is, and has always been, seen by Jews as very much a part of the Diaspora.

Some insight, we believe can be gained by examining the difference between 'non-violence' and 'not violent'. Non-violence is the epithet given to the political thought inspired by Gandhi and Martin Luther King among others in recent history. While 'non-violence' activism can contain an array of direct actions and coercive tactics such as hunger-strikes, acts of civil disobedience, sanctions, etc. it is the very opposite of the pro-violent politics of extremism that regards killing as necessary. Not-violent, on the other hand, is simply the absence of violence. Not-violence is simply one end of the full spectrum of asymmetric warfare. But one person's non-violence can be seen as another person's not-violence. Where does supposedly justifiable defensive violence cross over into terrorism? How is this ambiguity seen in a social context? For example, if Israel is responsible for the Middle East conflict then the BDS (Boycott, Divestment and Sanctions movement) is an example of a practical way to do something about it. However, many within the UK Jewish community are finding communal life increasingly difficult due to the BDS and have memories of the Nazis boycotting Jewish shops and goods in the past. Where is the line between non-violence and not-violence?[11]

---

[10] As seen not only by members of the French Jewish population who have recently emigrated to Israel because of increased Anti-Semitism but also by the majority of the 'forgotten' 850,000 Jews who did likewise because they were expelled from Muslim countries, with next to nothing, on the creation of the State of Israel.

[11] The difficulties in perceiving and determining these lines in the sand can be demonstrated by the following. During 2013 CST received reports of 465 potential incidents that, after investigation, were not included in the total of 529 anti-Semitic incidents. An example of one of the anti-Israel incidents (that was not recorded as Anti-Semitism even though it was using Anti-Semitic iconography of Jews being vermin and stealing from others) was a letter to a charity stating: "You are attempting to legitimize theft, you thieving c**ts! Give the Palestinians their land back. You f*****g parasites.". With respect to Western attitudes, the criticism of possible indiscriminate or non-proportional use of force is often couched in a language that may be interpreted as reminiscent of the Blood Libel. It may be that two thousand years of Christian ideology with respect to Jews is not easily removed. See for example, with respect to the chants of 'baby killer' https://en.wikipedia.org/wiki/Little_Saint_Hugh_of_Lincoln and https://www.facebook.com/israeladvocacymovement/videos/vb.829113587172986/928146197269724/?type=2&theater. The same can also be said of extremist Islamic religious ideology: see footage from a demonstration in London on 08 December 2017. Calls can be heard of 'Khaybar Khaybar, ya yahud, Jaish Muhammad, sa yahud' – Jews, remember Khaybar, the army of Muhammad is coming. The placards in this video are from the Palestine Solidarity Campaign (of whom, Jeremy Corbyn, Leader of the UK Labour Party is a patron) https://youtu.be/Zi2X19xaH0Y.

As Galam argues [17]: "The main feature of a passive supporter is that it is someone who would not oppose a terrorist move if it could, since it shares the terrorist goal, at least in part. Or it could also be because the supporter opposes the political power which is fighting terrorism. It is of importance to emphasize that most passive supporters are never confronted by any terrorist move. It might possibly be perceived as a neutral attitude towards the political content of a terrorist trend." It is significant that both the extreme Left and extremist Islamists – both ideologies that hope for some future 'perfect' state - criticize the USA and Israel and are viewed as the source of all the world ills[12]. It is also noteworthy that the USA and Israel were countries both born from the desire for freedom – from the moral perspective of Western, Judeo-Christian culture (as exemplified in the US Constitution and the desire by a significant part of the European and Middle Eastern Jewish population's wish for the freedom not to be persecuted[13][14]). Our main point therefore is that it is not at all clear what can constitute the numerical population value for passive support within a general population: 10%? 25%? A value may possibly be approximated by the collection of granular, individualized data.

One final issue with the Galam model is the calculation of the number of memes fed into the universal percolation formula. The meme table is not a scorecard - thus every meme has an equal weighting - and the model does not presently distinguish between the types of terrorist events that affect the socio-political geographical landscape. Lone wolves are treated the same as politically motivated attacks (the Madrid bombings, for example, were designed to cause a change in an election result). It is quite apparent, furthermore, that other groups besides jihadists (for example, anti-globalization terrorists) may use similar memetic methods[15] in perpetrating their agenda.

Our response therefore is twofold:
- Obtain additional data by which the percolation model can be further examined;
- Develop a fractal based dynamic meta model that describes spatio-temporal aspects of terrorist events (in a similar way that we undertook with attrition warfare see [4])

## 3.0 Summary

In this paper, we have initiated an attempt to develop and understand the physics that underlies terrorist events. Our main conclusion from this initial study is that the non-active permeation of terrorist ideologies (which we call 'passive support') through the social, political and psychological fabric of society not only contributes to the active participation of terrorists within society but also directly contributes to and increases the likelihood of the occurrence of terrorist events. However, from the perspective of our previous work on the physics of human systems [4], [5], [19] this current paper merely provides insight into the highest tier of the modeling environment (the 'strategic' or 'percolation' tier). In a further paper, we hope to demonstrate a more extensive Galam model - going some way towards addressing the questions raised above - expanding our study to non-European data together with constructing a model , at the middle, or 'operational' tier, that is, a fractal dynamic model. The final and most granular, behavioral, tier, the 'tactical' tier, will be developed as part of a further study. The advantage of constructing such a model hierarchy is that each level would mutually validate the other model levels.

---

What is significant to our study is that the incident was little reported in the UK national press, indicating that it was not newsworthy (the implication therefore is that no one cared i.e. we are observing an example of passive support and another case of casual anti-semitism from the general UK population).

[12] See for example Noam Chomsky's series of lectures e.g. https://youtu.be/lUQ_0MubbcM
[13] See for example https://en.wikipedia.org/wiki/Antisemitism_in_Europe
[14] See https://en.wikipedia.org/wiki/Category:Anti-Jewish_pogroms_by_Muslims
[15] Rapid mimetic communication characterizes social media usage. Social media, however, is further characterized by individuals ability to tune the content transmitted and viewed. The net result is that perspectives and opinions are even greater polarized. There is a suggestion that 21st century, terrorism is morphing. A unique combination of past doctrines, modern recruiting methods, advanced technology, and extremists' relentless desire to destroy all that is not within the purviews of Islam point to a further 'wave' of terrorism in the near future. More importantly, Islamic extremism does not appear to be the sole contributor to this possible new wave of terrorism [18].

**Figure 11. Meta-Model Hierarchy**

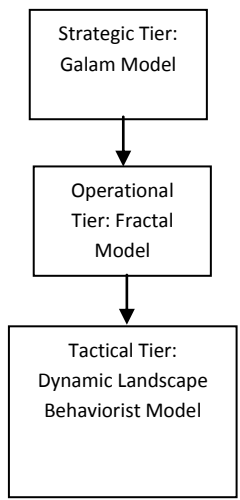

This further study will propose novel applications of complex adaptive systems theory to develop massive automated parallel processing systems which will enable the tracking and mapping of the dynamics of money laundering behaviors and derive their causal and identification factors in real time. The methodology of this 'operational' tier is similar to that proposed in our previous work on the Fractal nature of Warfare [3] and consists of undertaking a mathematical study of a dynamical system whose solutions are the non-linear equations that exhibit complex temporal and spatial behavior. The underlying concept behind such studies is that of the chaotic or strange attractor and the parameters which define the basin of attraction of the phenomenon under study. These attractors are different from simple attractors (point attractors) or attractors that decay into periodic states (limit cycles) in that they decay into a non-periodic and complex final state. Once this final state is identified, a number of measures can be taken in order to identify and quantify its nature. The methodology, therefore, consists of undertaking data analysis using high-performance computing systems to create algorithms that identify and predict criminal or terrorist behavior not merely by looking at previous data characteristics and pattern matching but rather by developing the 'physics' behind these networks and creating a multidimensional, dynamic 'landscape' or phase space that describes and characterizes this complex behavior. Building on the mathematical construct of this landscape we could then identify and examine criminal behavioral 'attractors' within the landscape. These attractors would include salient features from the demographic, political, financial and legal landscape. This research would constitute the first attempt to model these factors in this fashion in this type of context. For 4GW, data sources for both the USA and the rest of the world are available from the Global Terrorism Database[16]. This would allow an initial study at the operational level to be undertaken.

Lastly, the UK government has recently implemented the Mindspace program on how public behavior may be influenced using public policy[17]. This program demonstrates because how behavioral theory, together with collection of individualized data, could help achieve better outcomes for citizens, either by complementing more established policy tools, or by suggesting more innovative interventions. In doing so, this draws upon the most recent academic evidence, as well as exploring the wide range of existing good work in applying behavioral theory across the public sector. The behaviorist methodologies and data collected would be able to provide an initial framework for a tactical tier model[18].

---

[16] See http://www.start.umd.edu/gtd/
[17] https://www.instituteforgovernment.org.uk/sites/default/files/publications/MINDSPACE.pdf
[18] Examples of the practical implementation of Big Data collection, analysis and use of behavioural science on the back of the Mindspace program can be seen at http://www.thebehaviouralist.com